\documentclass[twocolumn,showpacs,prc]{revtex4}
\usepackage{dcolumn}
\usepackage{bm}

\begin{document}

\title{Thermal Time Scales in a Color Glass Condensate}
\author{V. Parihar and A. Widom}
\affiliation{Physics Department, Northeastern University, Boston MA 02115, USA}
\author{Y.N. Srivastava}
\affiliation{Physics Department \& INFN, University of Perugia, Perugia, IT}

\begin{abstract}

In a model of relativistic heavy ion collisions wherein 
the unconfined quark-gluon plasma is condensed into glass, 
we derive the Vogel-Fulcher-Tammann cooling law. This law 
is well known to hold true in condensed matter glasses.  
The high energy plasma is initially created in a very hot 
negative temperature state and cools down to the Hagedorn glass 
temperature at an ever decreasing rate. The cooling rate is largely 
determined by the QCD string tension derived from hadronic Regge 
trajectories. The ultimately slow relaxation time is a defining 
characteristic of a color glass condensate.

\end{abstract}

\maketitle

\section{Introduction \label{Intro}}
There has been considerable recent 
interest\cite{Maiani,Shuryak:2004,Shuryak:2002,Soff:2001,Teaney:2001} 
in  measurements of a possible quark-gluon plasma 
whose properties may be probed by relativistic 
heavy ion collisions (RHIC).  Central to the study 
of a possible unconfined color plasma  
is the time scale required to form this thermodynamic 
fluid phase. The central question concerns how the  
phase formation time compares with the collision time 
of the heavy ion probes. 

The apparent experimentally observed suppression of 
heavy quark particle production has led to the notion 
that thermal formation times may be longer than might 
be at first expected.  The notion of an unconfined 
color fluid plasma was, for {\em some} RHIC 
energies, replaced by the notion of a color glass 
condensate\cite{Gtulassy,Kharzeev,Iancu,Jalilian,Venugopalan,
Iancu:02,Venugopalan:97,Ayala} (CGC). In simple terms, thermal 
relaxation times in glasses are 
much longer than thermal relaxation times in fluids. If the thermal 
relaxation times were {\em not} much shorter than the RHIC 
collision times, then the observed heavy quark suppression 
becomes understandable.    
   
The thermal relaxation time in condensed matter glasses follows 
a well known and long established law due to 
Vogel\cite{Vogel:1921}, Fulcher\cite{Fulcher:1925} 
and Tammann\cite{Tammann:1926} (VFT). The thermal 
relaxation time depends on temperature according to the 
VFT rule\cite{Tanaka:2003}
\begin{equation}
\tau \approx \tau_\infty   \exp\left[\frac{\Phi }{k_B(T-T_g)} \right]
\ \ \ \ \ (T> T_g),
\label{intro1}
\end{equation}
wherein \begin{math} \Phi \end{math} is a thermal activation 
energy and \begin{math} T_g  \end{math} is a dynamical 
glass temperature. Our purpose is to point out that the 
VFT Eq.(\ref{intro1}) is expected to hold true for glass phases 
obtained from QCD inspired models. The derivation of Eq.(\ref{intro1})  
will be given and the method for determining the parameters will 
thereby be provided.

The notion that the perturbation theory QCD ``vacuum'' is in reality 
an excited state at a { \em negative} temperature\cite{pacetti} is 
introduced in Sec.\ref{pert}. In terms of the dynamic dielectric response function
\begin{math} \varepsilon \end{math}, the imaginary part is negative for 
the QCD ``vacuum'' and positive for QED vacuum. It is the negative 
temperature feature of the perturbation theory QCD vacuum that allows 
for the avoidance of Landau ghosts\cite{Landau:1955,siva}. The color 
screening \begin{math} \varepsilon \end{math} directly yields a QCD ``string 
potential'' between interacting quarks.
In Sec.\ref{strings} we discuss the QCD strings whose tension 
\begin{math} \sigma  \end{math} is empirically described by hadronic 
Regge trajectories; In Sec.\ref{Hag}, the dynamical glass  
temperature \begin{math} T_g  \end{math} will be explored via the 
Hagedorn\cite{Hagedorn:1965} entropy. We estimate a glass temperature 
and an activation energy, respectively, of 
\begin{equation}
k_BT_g=\sqrt{3\hbar c \sigma \over 4\pi }\approx 207\ MeV
\ \ {\rm and}\ \ \Phi \approx 725\ MeV.
\label{intro2}
\end{equation}
At its inception, the Hagedorn temperature was viewed as the largest possible 
temperature that could be achieved by smashing hadrons together at very high 
center of mass energy. Presently, the use of QCD perturbation theory for very high 
energies implies for short time scales unconfined color, i.e. almost free 
quarks and gluons. 

For high energy RHIC events, the almost free initially produced quarks and 
gluons constitute a very hot plasma at negative temperature. The plasma 
cools at first to even more negative temperatures reaching 
\begin{math} T\to -\infty  \end{math} and entering at 
\begin{math} T\to +\infty  \end{math}. In other words, it is the negative 
{\em inverse temperature} \begin{math} \beta \equiv (k_BT)^{-1}\end{math} 
(and {\em not} \begin{math} T  \end{math}) which passes through zero when you  
cool a system which starts at negative temperature. Now the plasma further 
quickly cools from \begin{math} \infty > T\gg T_g  \end{math} heading towards 
the Hagedorn temperature \begin{math} T_g \end{math} from above. But in accordance 
with VFT glass asymptotic Eq.(\ref{intro1}), the cooling relaxation time grows 
exponentially ever larger; 
\begin{equation}
\tau \to \infty \ \ \ {\rm as} 
\ \ \ T\to T_g+0^+.
\label{intro3}
\end{equation} 
The collision ends before the temperature can get below the Hagedorn 
temperature. The glass never fully hardens as discussed in the 
concluding Sec.\ref{conc}.

\section{Perturbation Theoretical QCD \label{pert}}

Recall the running coupling constant in QED, i.e. 
\begin{equation}
\alpha (Q^2)=\frac{e^2}{4\pi \hbar c \varepsilon (Q^2)}\  ,
\label{pert1}
\end{equation}
wherein the dielectric response of the vacuum obeys the 
dispersion relation 
\begin{equation}
\varepsilon (Q^2)=1-\frac{Q^2}{\pi }\int_0^\infty 
\left[\frac{{\Im}m\ \varepsilon(-\nu-i0^+)}{\nu+Q^2}\right]\frac{d\nu }{\nu }\ .
\label{pert2}
\end{equation}
For time-like wave vectors, the vacuum is dissipative
\begin{equation}
{\Im}m\ \varepsilon(-\nu-i0^+)  \ge  0 
 \ \ \  {\rm if} \ \ \ \nu=-Q^2>0.
\label{pert3}
\end{equation}
Among the processes contributing to the dissipative 
vacuum Eq.(\ref{pert3}) are the creation out of the vacuum of 
charged particle anti-particle pairs produced by incident 
electromagnetic radiation. From Eqs.(\ref{pert2}) and (\ref{pert3}), 
it follows that for some space-like \begin{math} Q^2>0  \end{math}, 
the real dielectric response will vanish. This constitutes the Landau 
ghost problem of QED. In particlular, the potential energy between 
two static charges, \begin{math} z_1e  \end{math} and 
\begin{math} z_2e  \end{math} is determined by the dielectric 
screening function  
\begin{math} e^2 \to e^2/\varepsilon(Q^2) \end{math}  
as discussed in standard works on quantum 
electrodynamics\cite{Berestetskii:1997}; 
\begin{eqnarray}
V(r) &=& \left[\frac{z_1z_2e^2}{4\pi r}\right] \chi(r),
\nonumber \\ 
\chi(r) &=& \frac{2}{\pi }\int_0^\infty \sin(Qr)\left[\frac{dQ}{Q\varepsilon (Q^2)}\right] .
\label{pert4}
\end{eqnarray}
Note that \begin{math}  \lim_{r \to \infty} \chi(r)=1  \end{math}.

For the QCD case, the running coupling constant 
\begin{equation}
\alpha_s (Q^2)=\frac{g^2}{4\pi \hbar c \varepsilon_s (Q^2)}\  ,
\label{pert5}
\end{equation}
wherein
\begin{equation}
\varepsilon_s (Q^2)=-\frac{Q^2}{\pi }\int_0^\infty 
\left[\frac{{\Im}m\ \varepsilon_s(-\nu-i0^+)}{\nu +Q^2}\right]\frac{d\nu}{\nu }\ .
\label{pert6}
\end{equation}
To one loop order in perturbation theory, one finds 
\begin{eqnarray}
\frac{{\Im m}\  \varepsilon_s(-\nu -i0^+)}{\pi }
&=& -\frac{g^2}{4\pi \hbar c}(\beta_0  + \ldots )
\nonumber  \\
\beta_0 &=& \frac{1}{4\pi }\left(\frac{11}{3}N_c-\frac{2}{3}n_f\right)
\label{pert7}
\end{eqnarray}
wherein \begin{math} N_c  \end{math} is the number of colors and 
\begin{math} n_f  \end{math} is the number of flavors. 

Note the condition 
\begin{equation}
{\Im m}\  \varepsilon_s(-\nu -i0^+) \le 0\ \ \ {\rm (color\ amplifier)},
\label{pertt8}
\end{equation}
which implies that the perturbation theory vacuum is in reality an excited 
QCD state which decays into the true vacuum\cite{pacetti}. Unlike the perturbation 
theory QED vacuum which requires external radiation energy to produce
particle pairs, the perturbation theory QCD vacuum {\em spontaneously} decays 
into the true vacuum radiating bursts of hadrons. The excited state QCD perturbation 
theory vacuum is similar to an excited amplifying LASER material 
with inverted excitation energy levels. The Laser material spontaneously 
decays into the true ground state radiating a photon pulse. 
Amplifying media may be described by a {\em negative} temperature. 
Initial high energy particles blast the true vacuum into a 
QCD perturbation theory vacuum at negative temperature. The resulting 
system then cools to the true vacuum radiating hadrons. When a very hot 
negative temperature cools down the temperature keeps going down. 
The temperature goes down through \begin{math} T=-\infty  \end{math} 
then appearing at \begin{math} T=+\infty  \end{math} further cooling 
through decreasing positive temperatures. From Eqs.(\ref{pert6}) and 
(\ref{pertt8}) it follows for space-like wave vectors that 
\begin{math} \varepsilon_s(Q^2)>0  \end{math} so there are no 
QCD Landau ghosts.

The quark potentials are described in terms of the color screening function 
\begin{eqnarray}
V_s(r) &=& \eta^{ab}t_{1a}t_{2b}\left[\frac{g^2}{4\pi r}\right] \chi_s(r),
\nonumber \\ 
\chi_s(r) &=& \frac{2}{\pi }\int_0^\infty \sin(Qr)
\left[\frac{dQ}{Q\varepsilon_s (Q^2)}\right] ,
\label{pert9}
\end{eqnarray}
where the color matrices for the two quarks are 
\begin{math} t_{1a}  \end{math} and \begin{math} t_{2b}  \end{math}
Taking two derivatives of Eq.(\ref{pert9}) with respect to 
\begin{math}  r  \end{math} yields  
\begin{equation}
\chi^{\prime \prime }_s(r)=-\frac{2}{\pi }\int_0^\infty \sin(Qr)
\left[ \frac{QdQ}{\varepsilon_s(Q^2)} \right].
\label{pert10}
\end{equation}
From the small wave number dependence of the color dielectric response,
\begin{eqnarray}
\frac{L^2}{2} &=& \lim_{Q\to 0} \frac{\varepsilon_s(Q^2)}{Q^2},
\nonumber \\ 
\frac{L^2}{2}  &=& -\frac{1}{\pi }
\int_0^\infty {\Im}m\ \varepsilon_s(-\nu -i0^+)\frac{d\nu }{\nu^2 }>0,
\label{pert11}
\end{eqnarray}
one finds via Eqs.(\ref{pert10}) and (\ref{pert11}) that 
\begin{eqnarray}
\lim_{r\to \infty}\chi^{\prime \prime }_s(r) &=& -\left(\frac{2}{L^2}\right) ,
\nonumber \\ 
\chi_s(r\to \infty) &=& -\left(\frac{r}{L}\right)^2 .
\label{pert12}
\end{eqnarray}
The central result of this section is a confining linear potential 
\begin{equation}
V_s(r) = - \eta^{ab}t_{1a}t_{2b}\sigma r\ \ \ {\rm as}\ r\to \infty ,
\label{pert13}
\end{equation} 
with a ``string tension'' 
\begin{equation}
\sigma =\left(\frac{g^2}{4\pi L^2}\right).
\label{pert14}
\end{equation} 
Let us now consider in more detail the nature of the QCD string.

\section{QCD Strings \label{strings}}

The QCD string may be physically pictured as follows\cite{yogi}: (i) Between 
two quarks with color charges \begin{math} t_{1a}g  \end{math} 
and \begin{math} t_{2b}g  \end{math} is a string. (ii) Inside the string 
is a gluon condensate electric field 
\begin{equation}
{\cal E}=\sqrt{\eta^{ab}\left<
{\bf E}_a \cdot {\bf E}_b - {\bf B}_a \cdot {\bf B}_b
\right>}=\left(\frac{g}{4\pi L^2}\right).
\label{strings1}
\end{equation}
(iii) The tension in the string is   
\begin{equation}
\sigma =g{\cal E}.
\label{strings2}
\end{equation}
(iv) Let us suppose two virtually zero mass quarks move 
with light speed \begin{math} c \end{math} on the ends 
of a linear string which extends along the straight line segment 
\begin{math}  -a< r <a  \end{math}. In rigid body rotation 
\begin{eqnarray}
\frac{|{\bf v}|}{c} &=& \frac{|r|}{a}\ \ \ ({\rm speed }),
\nonumber \\
Mc^2 &=& \int_{-a}^a \frac{\sigma dr}{\sqrt{1-|{\bf v}/c|^2}},
=\pi \sigma a ,
\nonumber \\ 
J &=& \frac{1}{c^2}
\int_{-a}^a \frac{\sigma r|{\bf v}|dr}{\sqrt{1-|{\bf v}/c|^2}}
=\left(\frac{\pi \sigma a^2}{2c}\right),
\nonumber \\ 
J &=& \frac{M^2c^3}{2\pi \sigma }\ \ \ ({\rm classical}).
\label{strings3}
\end{eqnarray}
The quantum relationship between angular momentum 
\begin{math} J \end{math} and mass \begin{math} M§  \end{math} 
is taken to be 
\begin{equation}
J=\hbar \alpha_0 +\frac{M^2c^3}{2\pi \sigma }.
\label{strings4}
\end{equation}
From experimental linear Regge trajectories 
\begin{equation}
\hbar c \sigma \approx 0.18\ GeV^2\ .
\label{strings5}
\end{equation}

The question arises as to the nature of quark-anti quark pair creation,  i.e. jet 
production,  in a scattering experiment wherein the incoming particles supply the 
energy to create the gluon condensate within the string as well as supplying 
the quark pair energy (\begin{math} mc^2  \end{math} per quark).  
Quark motion along the string obeys Newton's law for the rate of change
of momentum 
\begin{equation}
\frac{dp}{dt}=g{\cal E}=\sigma .
\label{strings8}
\end{equation}
The quark energy-momentum relation in a 1+1 dimensional QCD string 
follows from the partitioned Hamiltonian matrix 
\begin{equation}
H=
\pmatrix{ cp & mc^2 \cr
mc^2 & -cp},
\label{strings9}
\end{equation} 
where (from the quark-pair creation viewpoint) the transition rate per unit time to 
create a quark is determined by Fermi's golden rule 
\begin{equation}
\bar{\nu }=\frac{2\pi}{\hbar}|mc^2|^2\delta (2cp).
\label{strings10}
\end{equation}
From the viewpoint of QCD string fragmentation,  the probability that the 
string  dissociates may be written
\begin{eqnarray}
d^2P &=& \frac{dp dr}{2\pi \hbar}\exp\left[-\int \bar{\nu} dt \right],
\nonumber \\ 
\frac{d^2P}{drdt} &=& 
\frac{dp}{2\pi \hbar dt}\exp\left[-\int \bar{\nu} \frac{dt}{dp}dp \right], 
\nonumber \\ 
\frac{d^2P}{drdt} &=& 
\frac{g{\cal E}}{2\pi \hbar }\exp\left[-\int \bar{\nu} \frac{dp}{g{\cal E}} \right],
\label{strings11}
\end{eqnarray}
where Eq.(\ref{strings8}) has been invoked.
The transition rate per unit time per unit string length for fragmentation 
follows from Eqs.(\ref{strings10}) and (\ref{strings11}); It is  
\begin{equation}
\gamma = \frac{g{\cal E}}{2\pi \hbar }e^{-m^2c^3/\hbar g{\cal E}}
= \frac{\sigma }{2\pi \hbar } 
e^{-m^2c^3/\hbar \sigma} .    
\label{strings12}
\end{equation}
From the above results, it appears that strings connecting high 
mass quarks are less likely to fragment than strings connecting low 
mass quarks. The QCD system starts out very hot so that  critical 
glass temperatures are approached from above. It is here that the 
notion of a string glassy state appears most useful.  

\section{Hagedorn Glass Temperature \label{Hag}}

We have shown in Eq.(\ref{strings3}) for a classical QCD string model 
that the energy and angular momentum are related by  
\begin{equation}
\frac{E^2}{2\pi c\sigma }=J.
\label{hag1}
\end{equation}
In the quantum theory of the Boson (gluon) string, \begin{math} J  \end{math} 
has the spectrum
\begin{equation}
J=\hbar N\ \ \ {\rm where}
\ \ \ N=0,1,2,3,\ldots\ .
\label{hag2}
\end{equation}
If there are \begin{math} n_{k,j}=0,1,2,3, ....  \end{math} string excitations 
with polarization \begin{math} j \end{math} each carrying 
angular momentum \begin{math} \hbar k \end{math}  
with  \begin{math} k=1,2,3, ....  \end{math}, then  
\begin{equation}
N=\sum_{j=1}^2 \sum_{k=1}^\infty k n_{kj}.
\label{hag3}
\end{equation}
Eq.(\ref{hag3}) gives rise to a statistical mechanical entropy problem 
which involves analytical number theory. How many different ways 
\begin{math} \Omega  \end{math} can you 
form the integer \begin{math} N  \end{math} out of the smaller 
integers \begin{math} \{n_{kj}\}  \end{math}?    
From the asymptotic solution\cite{Hardy:1918} of an earlier and similar 
partitioning problem, the solution to the QCD excitation string 
partitioning problem was found\cite{Huang:1970}. In the large 
\begin{math} N\to \infty  \end{math}  limit, 
\begin{equation}
\ln \Omega(N)\approx 2\pi \sqrt{2N\over 3}+\ln\left[(6N)^{-7/4}\sqrt{3}\right]
\label{hag4}
\end{equation}
From the general definition of degeneracy and entropy, 
\begin{equation}
S=k_B \ln \Omega ,
\label{hag5}
\end{equation}
along with Eqs.(\ref{hag1}), (\ref{hag2}) and (\ref{hag4}) we find the 
Hagedorn string entropy 
\begin{eqnarray}
\frac{S}{k_B}&=&\left(\frac{E}{k_BT_g}\right)-
\frac{7}{2}\ln\left[\frac{E}{k_BT_g}\right]+\frac{\tilde{S}}{k_B},
\nonumber \\
\frac{\tilde{S}}{k_B} &=& 
\frac{1}{2}\left[\ln(3)+7\ln\left(\frac{2\pi }{3}\right)\right]. 
\label{hag6}
\end{eqnarray}
The Hagedorn glass temperature is 
\begin{equation}
k_BT_g=\sqrt{3\hbar c \sigma \over 4\pi }\approx 207\ MeV 
\label{hag7}
\end{equation}
where Eq.(\ref{strings5}) has been invoked.
The thermal equations of state for a hot string now follow from 
\begin{eqnarray}
\frac{1}{T} &=& \frac{\partial S}{\partial E}
\nonumber \\ 
\frac{1}{T} &=& \frac{1}{T_g}-\frac{7k_B}{2E}.
\label{hag8}
\end{eqnarray}
Thus, the entropy as a function of temperature reads,  
\begin{eqnarray}
\frac{S}{k_B} &=& \frac{7}{2}\left(\frac{T_g}{T-T_g}\right)
\nonumber \\ 
&-&\left(\frac{7}{2}\right)\ln\left[\frac{T}{(T-T_g)}\right]+
S_\infty \ \ \ \ \ (T>T_g),
\nonumber \\
S_\infty &=& \lim_{T\to \infty}S(T)
\nonumber \\
\frac{S_\infty}{k_B} &=&
\frac{7}{2}\left[1+\ln\left(\frac{2\pi }{3}\right)
-\ln\left(\frac{7}{2}\right)\right]+\frac{1}{2}\ln 3.
\label{hag9}
\end{eqnarray}

The role of the entropy in determining transition rates follows from the 
rule of averaging over initial states and summing over final states. Since 
the ratio of the number of final states to the number of initial states is 
given by 
\begin{equation}
\frac{\Omega_f}{\Omega_i}=\exp\left[\frac{S_f-S_i}{k_B}\right],
\label{hag10}
\end{equation}
the transition rates may be phase space dominated by the exponential 
entropy factors. In particular, the thermal relaxation times for very 
hot \begin{math} T\to \infty  \end{math} 
color unconfined states is related to the thermal relation times for
finite temperature \begin{math} T \end{math} unconfined states via 
\begin{eqnarray}
\tau_\infty &=& \tau \exp\left[\frac{S_\infty -S}{k_B}\right] 
\ \ \ (T>T_g), 
\nonumber \\ 
\tau &=& \tau_\infty \left(\frac{T-T_g}{T} \right)^{7/2}
\exp\left[{\Phi \over k_B(T-T_g)}\right]
\label{hag11}
\end{eqnarray}
Eq.(\ref{hag11}) is consistent with the VFT asymptotic Eq.(\ref{intro1}) 
with 
\begin{equation}
\Phi=\frac{7k_BT_g}{2}\approx 725\ MeV.
\label{hag12}
\end{equation} 
Eq.(\ref{hag11}) is the central result of this work.
\medskip

\section{Conclusion \label{conc}}

Perturbative QCD, presumed valid at high energies or for very 
short times, contains free quarks and gluons as manifested 
through jets and leads naturally towards the possibilty of a quark-gluon 
plasma phase in high energy heavy ion collisions(RHIC). 
Our analysis of quark-gluon plasma formation in heavy ion collisions 
devolves around the fact that the perturbative QCD vacuum is unstable. 
We have shown that in reality it is an excited state at a negative 
temperature. The negative temperature feature of the perturbative QCD 
vacuum is responsible for the avoidance of the Landau ghost in QCD
\cite{pacetti,siva}. 

On the other hand, certain results from RHIC experiments regarding 
heavy flavor production and jets led to the conjecture that the thermal 
quark-gluon plasma formation times may be longer than the collision times
and hence to the notion of a new state of matter: a color glass condensate. 
For a glassy state, thermal relaxation times are long and a collision ends 
before the glass fully hardens. We have demonstrated here that in the hadronic 
QCD string model the unconfined quark-gluon plasma is indeed condensed into 
glass. Our computation of the thermal relaxation times for unconfined
color states invokes the fact that the hot plasma begins at a negative 
temperature (as for a laser). We have derived the VFT cooling law for glass 
\begin{math} Eq.(38)\end{math} for unconfined quark-gluon plasma with 
a glass transition temperature estimated to be \begin{math} 207\ MeV \end{math} 
and an activation energy of  \begin{math} 725\ MeV  \end{math}. 
We have interpreted the glass transition temperature as the Hagedorn 
temperature (approached from above) at which collisions cease as a result 
of rapidly growing thermal relaxation times. It appears quite satisfactory
that asymptotic freedom from perturbative QCD in concert with a hadronic string
with its condensed color electric fields and a highly degenerate Regge 
spectrum, suffices to produce the VFT cooling law. It lends strong 
support to the existence of a color glass state of matter.

\end{document}